\documentclass[prl,superscriptaddress,showpacs,amssymb,amsmath,
amsfonts,aps,twocolumn,secnumarabic,floatfix]{revtex4}
\usepackage{graphics}
\usepackage{epsfig}
\begin{document} 
\bibliographystyle{try} 
 
\topmargin 0.1cm 
 
 \title{The space-time structure of hard scattering processes}

\newcommand*{\SACLAY }{ CEA-Saclay, Service de Physique Nucl\'eaire, F91191 Gif-sur-Yvette, Cedex, France} 
\affiliation{\SACLAY } 

\newcommand*{\JLAB }{ Thomas Jefferson National Accelerator Facility, Newport News, Virginia 23606} 
\affiliation{\JLAB } 

\author{J.M.~Laget}
     \affiliation{\SACLAY}
     \affiliation{\JLAB}

\date{\today} 
 
\begin{abstract}

Recent studies of exclusive electroproduction of vector mesons at JLab make it possible for the first time to play with two independent hard scales: the virtuality $Q^2$ of the photon, which sets the observation scale, and the momentum transfer $t$ to the hadronic system, which sets the interaction scale. They reinforce the description of hard scattering processes in terms of few effective degrees of freedom relevant to the Jlab-Hermes energy range. 
\end{abstract} 
 
 
\maketitle 

\begin{figure}[h]
\centerline{\epsfxsize=3.in\epsfbox{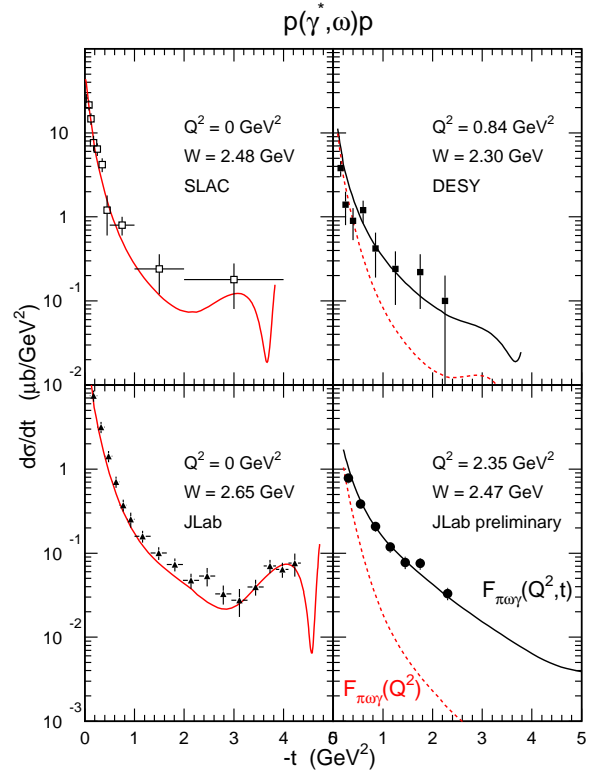}}   
\caption{(Color on line) The cross section of the production of $\omega$ mesons by real (left) and virtual photons (right).
\label{dsdt_ome}}
\end{figure}

The study of exclusive electro-production of $\omega$ mesons, recently completed at JLab~\cite{Mo03}, provides us with an original insight on the space time structure of hard scattering processes between the constituents of hadrons. The data speak for themselves in Figure~\ref{dsdt_ome}. The high intensity of the CEBAF beam, combined with the large acceptance of CLAS, allowed us to perform measurement with an unprecedented accuracy: the two top panels show previous data, recorded 30 years ago or so with real photons at SLAC~\cite{Ba71} or virtual photons at DESY~\cite{Jo77}, while the two bottom panels show the JLab data~\cite{Mo03,Ba03}. The extension to higher virtuality $Q^2$ of the photon reveals the underlying reaction mechanisms. At low momentum transfer $-t$ (small angle), the variation of the cross section with $Q^2$ (from left to right panels) falls down as the electromagnetic form factor of the pion, the exchange of which dominates the $\omega$ channel~\cite{La00} in the JLab energy range. At large $-t$ (large angle) on the contrary, the cross section stays almost flat and points toward the coupling of the virtual photon to point-like objects.


At large momentum transfers (large angles) the impact parameter ($b\sim 1/\sqrt{-t}$) is small enough to force the partons to exchange the minimum number of gluons before they recombine into the final particles. These hard scattering processes are at the origin of the scaling rules~\cite{Le80}, which have been verified in many reactions: around 90$^{\circ}$, the cross section behaves as $s^{N-2}$, being $s=W^2$ the total available energy squared and $N$ the number of active constituents. However, a quantitative understanding of experimental cross sections has been difficult to achieve. In the simplest case, Compton scattering, perturbative calculations (see e.g.~\cite{Va97}) fall short by an order of magnitude for the cross section and predict spin transfer coefficients with a sign opposite to experiment~\cite{Wo03}. One is forced to rely on models based on effective partonic degrees of freedom relevant to the scale of observation, either Generalized Parton Distributions (GPDs)~\cite{Di99} or dressed quarks and gluons~\cite{La00, Ca02, Ca03}.

The photo-production of $\phi$ meson, which is dominantly made of a pair of strange quark-antiquark, selects two gluons exchange mechanisms~\cite{La00}. A fair agreement with the experiment~\cite{An00} is achieved when a dressed gluon propagator computed on Lattice and a correlated quark wave function of the proton are used~\cite{Ca02}.

In the photo-production of $\rho$ and $\omega$ mesons, light quark interchange processes are not forbidden and contribute in addition to two-gluon exchange. A fair agreement with the experiments~\cite{Ba01, Ba03} is achieved when saturating Regge trajectories~\cite{Ser94, Gui97} are used for the propagators of the various exchanged mesons~\cite{Ca02}. This is an economical way to deal with hard scattering mechanisms since the saturation of the Regge trajectories (approaching $-1$ when $-t \rightarrow \infty$)  is closely related to the one-gluon exchange interaction between quarks~\cite{Ser94}. The $\omega$ meson production channel is particularly instructive in this respect since pion exchange dominates the cross section. As can be seen in the left panels of Figure~\ref{dsdt_ome}, the agreement with the experiments is excellent (the rise and the node at the highest $-t$ are due to the exchange  of the $u$-channel nucleon non-degenerated Regge trajectory).

Real Compton Scattering can also be described in this approach. In the JLab energy range (4 to 6 GeV) a real photon fluctuates into vectors mesons, since they have the same quantum numbers, over a distance commensurate to the size of the nucleon. Real Compton Scattering and vector meson production observables are therefore related. Not only the differential cross section~\cite{Ca02} but also the spin transfer coefficients~\cite{Ca03} match the values recently determined at JLab~\cite{Wo03}.


\begin{figure}[h]
\centerline{\epsfxsize=3.5in\epsfbox{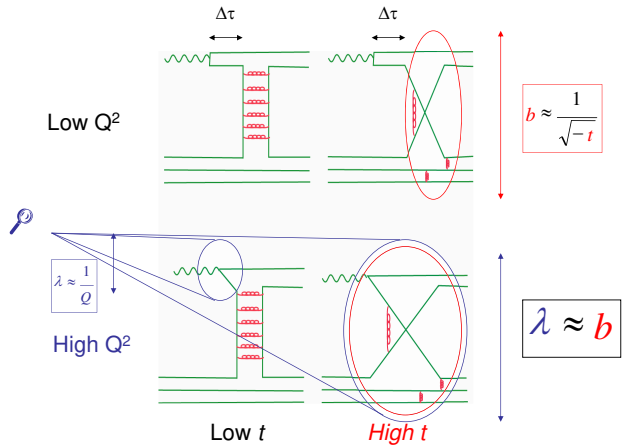}}   
\caption{(Color on line) A schematic view of the evolution of hard scattering mechanisms.
\label{space_time}}
\end{figure}

When the incoming photon becomes virtual (as it is exchanged between the scattered electron and the hadrons) two things happen. On the one hand, the lifetime $\Delta \tau=2\nu/(Q^2+m_V^2)$ of its hadronic component decreases: its coupling becomes more point-like. On the other hand, the transverse wave length ($\lambda \sim 1/Q$) of the photon decreases: it probes processes which occur at shorter and shorter distances. This is schematically depicted in Figure~\ref{space_time}. When both the virtuality $Q^2$ of the photon and the momentum transfer $-t$ are small (top left panel), the photon behaves as a beam of vector mesons which passes far away the nucleon target (large impact parameter $b$): the partons which may be exchanged have enough time to interact which each other and build the various mesons, the exchange of which drives the cross section. At high $-t$ (top right), the small impact parameter $b$ is commensurate to the hadronization length of the partons which must be absorbed or recombined into the final particles, within the interaction volume of radius $b$, before they hadronize. In other words, the two partons, which are exchanged between the meson and the nucleon, have just the time to exchange one gluon.  When $Q^2$ increases, the resolving power of the photon increases and allows it to resolve the structure of the exchanged quanta. When $-t$ is small (bottom left), it ``sees'' the partons inside the pion which is exchanged between the distant meson and nucleon. When $-t$ is large (bottom right), its wave length $\lambda$ becomes comparable to the impact parameter $b$: the virtual photon ``sees'' the partons which are exchanged during the hard scattering.

More quantitatively, the expressions of the various amplitudes, together with the corresponding coupling constants, are given in refs.~\cite{La00,Ca02,Ca03}. The $Q^2$ dependency is already built in the two-gluon as well as the $f_2$ meson exchange amplitudes. It happens~\cite{La95} to lead to the correct dependency as function of $Q^2$ and $-t$ (at least up to $-t\sim 1.5$~GeV$^2$)  in the $\phi$ meson electro-production sector~\cite{Dix77}, which emphasises two-gluon exchange. When the pion electromagnetic form factor
\begin{equation}
F_{\pi \omega \gamma}(Q^2)= \frac{1}{1+\frac{Q^2}{\Lambda_{0}^2}}
\end{equation}
with $\Lambda_{0}^2= 0.462$~GeV$^2$, is introduced at the $\pi \omega \gamma$ vertex of the $\pi$-exchange amplitude which reproduces real photon data, one obtains the dashed curves in the right panels of Figure~\ref{dsdt_ome}. They reproduce the evolution of the virtual photon cross section at low momentum transfer, but fall short by more than an order of magnitude at large momentum transfer. The agreement is restored (full curves) when a dependence against $-t$ is given to the pion form factor. It is natural to relate it to the way the pion saturating Regge trajectory, $\alpha_{\pi}(t)$~\cite{Gui97}, approaches its asymptote $-1$:
\begin{equation}
F_{\pi \omega \gamma}(Q^2,t)= \frac{1}{1+\frac{Q^2}{\Lambda_{\pi}^2(t)}}
\label{ff1}
\end{equation}
with
\begin{equation}
\Lambda_{\pi}^2(t)= \Lambda_{0}^2 \times \left(\frac{1+\alpha_{\pi}(0)}{1+\alpha_{\pi}(t)}\right)^2
\label{ff2}
\end{equation}
When $t\rightarrow -\infty$, $\alpha_{\pi}(t)\rightarrow -1$, and $F_{\pi \omega \gamma}(Q^2,t)$ becomes independent of $Q^2$ at large $-t$.

Such an ansatz links the evolution of $F_{\pi \omega \gamma}$, from the coupling to a full fledged pion toward the coupling to a point-like parton, with the underlying hard mechanism which dominates the cross section near 90$^{\circ}$, the exchange of two quarks which interact by exchanging the minimum number of gluons. While it is given to us by the experiment, it provides us with a quantity to be compared to a more fundamental theory, as Lattice Gauge calculations for instance. But also it provides us with links with other analysis. For instance, the effective radius of the partons which the virtual photon couples to at $-t\sim 2.5$~GeV$^2$ is about:
\begin{equation}
\sqrt{<r^2>} \sim \frac{\sqrt{6}}{\Lambda_{\pi}(-2.5)}\sim 0.15 \; \mathrm{fm}
\end{equation}  
very close to the value deduced from a recent analysis of the moments of the response functions of the nucleon in the JLab momentum range~\cite{Os03}.

\begin{figure}[h]
\centerline{\epsfxsize=2.9in\epsfbox{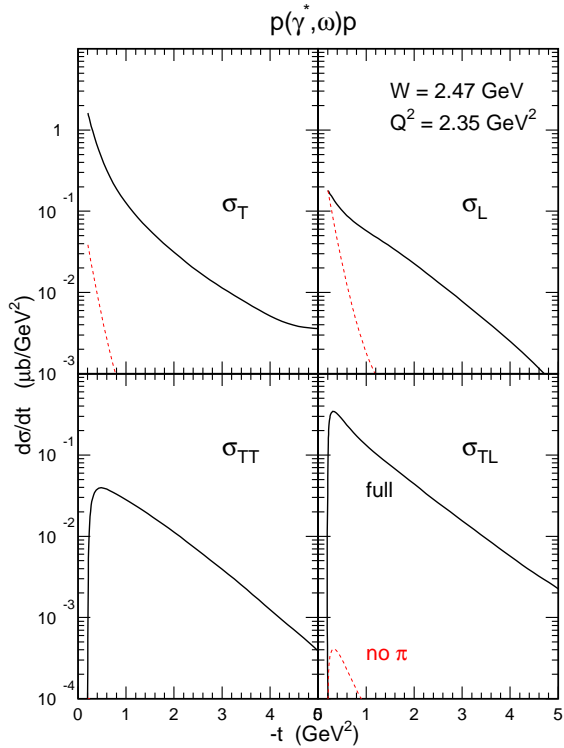}}   
\caption{(Color on line) The four parts of the cross section of the electro-production of $\omega$ mesons are plotted against $-t$. The dashed curves include the contributions of the two gluons and $f_2$ meson exchange amplitudes only. The full curves include also the contribution of the $\pi$ meson exchange.
\label{response_ome}}
\end{figure}

The $\omega$ electroproduction channel is dominated by the transverse part of the hadronic current. Fig.~\ref{response_ome} shows the four parts of the cross section, which I define as follows:
\begin{eqnarray}
\frac{d\sigma}{dE_e d\Omega_e dt}= \Gamma_v \times \left( \frac{d\sigma_T}{dt} +\epsilon \frac{d\sigma_L}{dt} + \epsilon \cos 2\Phi \frac{d\sigma_{TT}}{dt}
\right. \nonumber \\ \left.
 -\sqrt{\epsilon(1 + \epsilon)} \cos\Phi\frac{d\sigma_{TL}}{dt} \right)
\end{eqnarray}
where $\Gamma_v$, $\epsilon$ and $\Phi$ are respectively the flux of the virtual photon, its polarization and the angle between the hadronic plane and the leptonic plane (see e.g.~\cite{La94}). Two gluon- as well as $f_2$ meson- exchanges conserve helicity, contribute by the same amount to the Transverse ($\sigma_T$)and the Longitudinal ($\sigma_L$)parts of the cross section, and have a vanishing contribution to the Transverse-Transverse ($\sigma_{TT}$)and Transverse-Longitudinal ($\sigma_{TL}$) parts. On the contrary, pion exchange dominates the Transverse part and contributes little to the Longitudinal part (at least at low $-t$), but induces large interference cross sections. This prevents the identification of the Longitudinal and Transverse parts of the cross section from the decay angular distribution of the $\omega$ meson, assuming $s$-Channel Helicity Conservation (SCHC).

\begin{figure}[h]
\centerline{\epsfxsize=2.9in\epsfbox{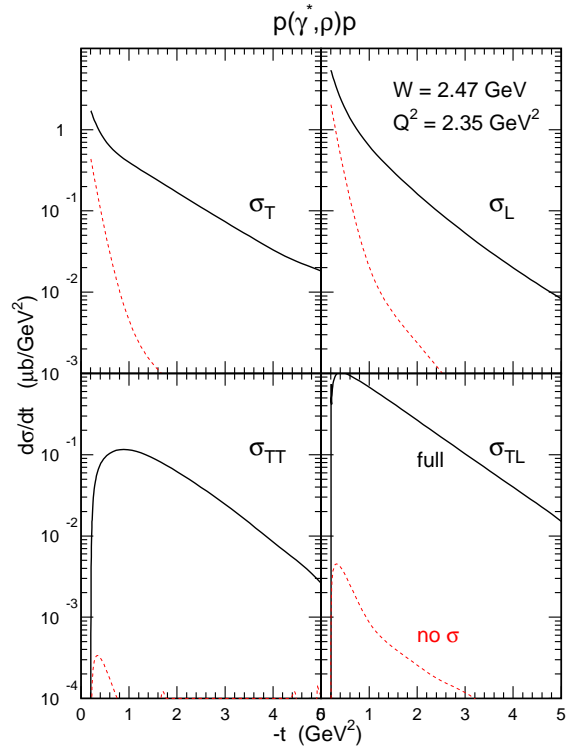}}   
\caption{(Color on line) The four parts of the cross section of the electro-production of $\rho$ mesons are plotted against $-t$. The dashed curves include the contributions of the two gluons and $f_2$ meson exchange amplitudes only. The full curves include also the contribution of the $\sigma$ meson exchange.
\label{response_rho}}
\end{figure}

The situation is different in the $\rho$ meson electro-production channel, which is dominated by the exchange of two gluons and $f_2$ meson (Fig.~\ref{response_rho}). The $\pi$ exchange contribution is vanishing, but the $\sigma$ meson exchange contributes. I use a dipole electromagnetic form factor $F_{\sigma \rho \gamma}$ with a cut-off mass $\Lambda_{0}= 1$~GeV$^2$, instead of the monopole with  $\Lambda_{0}= 0.462$~GeV$^2$ of~\cite{Ca03}.  Up to $Q^2 \sim 1.5$~GeV$^2$, the two choices are equivalent, but the dipole form factor agrees  better with the data at higher $Q^2$. I also let it depend on the momentum transfer $-t$ according to eqs.~\ref{ff1} and~\ref{ff2} where I use the saturating Regge trajectory of the $\sigma$ meson which leads to a good account of the $\rho$ meson photo-production at large $-t$~\cite{Ba01}. At low $-t$ (up to $\sim 1$~GeV$^2$), the smallness of the two interference response functions $\sigma_{TT}$ and $\sigma_{LT}$ indicates that SCHC is satisfied and that the Transverse and the Longitudinal parts can be determined from the analysis of the decay angular distribution of the $\rho$ meson. 

\begin{figure}[h]
\centerline{\epsfxsize=3.5in\epsfbox{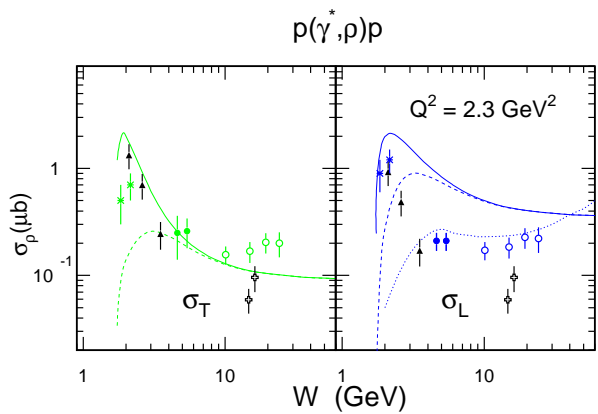}}   
\caption{(Color on line) The Transverse (left) and Longitudinal (right) cross sections of the electro-production of $\rho$ mesons at $Q^2=$ 2.3~GeV$^2$ are plotted against $W$. The dashed curves utilize $F_{\sigma \rho \gamma}(Q^2)$, while the full curves utilize $F_{\sigma \rho \gamma}(Q^2,t)$. The dotted curve is a GPDs prediction~\cite{Va99}. Star: Jlab preliminary~\cite{Cy03}; Filled triangles: Cornell~\cite{Ca81}; Filled circles: Hermes~\cite{He00}; Open circles: Fermilab~\cite{Fe97}; Open crosses: NMC~\cite{NMC}.
\label{rho_tot}}
\end{figure}

This study is under progress at JLab, and so far only the integrated Transverse and Longitudinal cross section have been extracted at JLab~\cite{Cy03} and before~\cite{He00,Fe97,Ca81,NMC}. Fig.~\ref{rho_tot} compares the prediction of the model to these data at $Q^2= 2.3$~GeV$^2$. The effect of the $-t$ dependency of the electromagnetic form factor manifests itself at low energy, since the minimum value $|t_{min}|$ of the momentum transfer, which is allowed by the kinematics, increases when the energy $W$ decreases. The cross section becomes more sensitive to the $-t$ dependency of the form factor. The Transverse cross section is well reproduced, but the Longitudinal cross section is over predicted by the model. At low $X < 10^{-2}$ this discrepancy has been resolved by fully taking into account~\cite{WFXX} the momentum dependency of the vector meson wave function. Whether this applies to larger $X\sim 0.3$ is still an open question.

An alternative approach of the Longitudinal part of the meson electroproduction cross section relies on GPDs. At high enough $Q^2$ the leading longitudinal amplitude factorizes into the perturbative production of a meson on a current quark and  GPDs which hide the complex non perturbative aspect of the nucleon target. The application~\cite{Va99} of the GPDs formalism down to low $Q^2\sim$~2.3 GeV$^2$ reproduces the Fermilab and Hermes data but falls short at lower energies, where the exchange of Regge trajectories cannot be neglected.  It turns out that such a factorized amplitude does not dominate the transverse amplitude. It is more sensitive to higher order mechanisms, which are more economically described in terms of a few effective degrees of freedom: dressed parton propagators, saturating Regge trajectories, electromagnetic form factors of off-shell meson. The success of this description in several channels is a strong hint that they are the relevant degrees of freedom in the JLab-Hermes energy range. In addition, they provide us with a link with more fundamental approaches of non perturbative QCD: ab initio Lattice Gauge calculations or potential models. 

In summary, the recent study of the electro-production of vector mesons at large momentum transfer has addressed a question which was posed but left unanswered for the past ten years. For the first time it has been possible to play independently with two hard scales: the virtuality $Q^2$ of the photon, which sets the observation scale, and the momentum transfer $t$ to the hadronic system, which sets the interaction scale. This has placed on solid ground the description of hard scatterings in terms of a few effective degrees of freedom, which my collaborators and I have developed over the past ten years or so. The determination of the dependency against the momentum transfer $t$ of the Longitudinal and the Transverse parts of the various meson electroproduction channels must be actively pursued in the present JLab energy range.  It will greatly benefit of its energy upgrade to 12 GeV, where one may expect that GPDs will become the relevant underlying degrees of freedom.

I acknowledge the warm hospitality  at JLab where this paper was completed.

\end{document}